\documentclass{article}
\usepackage{spconf,amsmath,graphicx}
\usepackage{cite}
\usepackage{amsmath,amssymb,amsfonts}
\usepackage{enumitem}
\usepackage{bm}
\usepackage{algorithmic}
\usepackage{graphicx}
\usepackage{textcomp}
\usepackage{xcolor}
\usepackage{booktabs}
\usepackage[hidelinks]{hyperref}
\usepackage{multirow}
\usepackage{tabularx}
\usepackage[ruled,vlined]{algorithm2e}
\DeclareMathOperator*{\argmax}{argmax}

\newcommand{\x}{\boldsymbol{x}}
\newcommand{\y}{\boldsymbol{y}}

\newcommand{\m}{\boldsymbol{m}}

\newcommand{\V}{\boldsymbol{V}}

\newcommand{\mc}{\mathcal{C}}
\def\x{{\mathbf x}}

\title{Self-Supervised Metric Learning With Graph Clustering \\  For Speaker Diarization}
\name{Prachi Singh \thanks{This work was supported by the grants from  the British Telecom Research Center.} and Sriram Ganapathy}
\address{Learning and Extraction of Acoustic Patterns (LEAP) Lab,\\
	Electrical Engineering, Indian Institute of Science, Bangalore, India.\\
	\texttt{prachisingh@iisc.ac.in, sriramg@iisc.ac.in}}
\begin{document}
	\ninept
	\maketitle
	\begin{abstract}
 In this paper, we propose a novel algorithm for speaker diarization using metric learning for graph based clustering. The graph clustering algorithms use an  adjacency matrix consisting of similarity scores. These scores are computed between speaker embeddings extracted from pairs of audio segments within the given recording. In this paper, we propose an approach that jointly learns the speaker embeddings and the similarity metric using principles of self-supervised learning. The metric learning network implements a neural model of the probabilistic linear discriminant analysis (PLDA).  The  self-supervision is derived from the pseudo labels obtained from a previous iteration of clustering. The entire model of representation learning and metric learning is  trained with a binary cross entropy loss. By combining the self-supervision based metric learning along with the graph-based clustering algorithm, we achieve significant relative improvements of $60\%$ and $7\%$  over the x-vector PLDA agglomerative hierarchical clustering (AHC) approach on AMI and the DIHARD datasets respectively in terms of diarization error rates (DER).
	\end{abstract}
	\section{Introduction}
	
	Speaker diarization, the problem of unsupervised temporal sequence segmentation into speaker specific regions, is one of first processing steps in the conversational analysis of multi-talker audio. The performance of a speaker diarization system is  adversely influenced by factors like short speaker turns, overlaps between multiple speakers,  far-field effects in audio recording and environmental artifacts.  The DIHARD evaluations explored various challenging environments for bench-marking diarization performance \cite{ryant2018first,ryant2019second,ryant2020dihard}.   
	
	In the past decade, the dominant approach to speaker diarization involved a two step process. The first step consists of deriving embeddings from relatively short windowed segments of speech (typically $1-2$s of audio) while the second step involves the  clustering of the embeddings. The early breakthroughs were reported with the  i-vector embedding extraction  using  unsupervised factor analysis modeling \cite{sell2014speaker}. The availability of large amounts of speaker supervised speech recordings along with the advancements in deep learning have propelled the use of deep neural network (DNN) based embedding extractors like the x-vectors \cite{snyder2018x}. 
	
	In terms of clustering, the common approach to speaker diarization is the bottom-up clustering approach where the goal is to successively merge to achieve a one-to-one correspondence between the ground-truth speakers and clusters~\cite{anguera2006robust}.  The most popular approach is the agglomerative hierarchical clustering (AHC) \cite{day1984efficient}.  The inputs to the clustering algorithms commonly employ pre-processing techniques on the embeddings like length normalization \cite{garcia2011analysis},  principal component analysis (PCA) \cite{zhu2016online} and   PLDA based affinity matrix computation ~\cite{sell2014speaker}. Another common approach to clustering is the spectral clustering approach~\cite{ning2006spectral}.  In most of these approaches, the affinity matrix computation and clustering are performed as two independent steps with different cost functions. 
	
	A neural diarization approach (termed as end-to-end (EEND))~\cite{shinji2019ASRU, shinji2020Interspeech}, proposed recently, explores  transformer models for speaker diarization. The key successes reported have been for recordings with $2-3$ speakers. However, training the EEND system with more speakers is challenging because of permutation-invariant loss computation. Further, the performance on recordings with more than $4$ speakers does not improve over the clustering based approaches~\cite{leapDIHARD3}. 
	
	The self-supervision in diarization  can provide effective representations for downstream tasks without requiring the ground-truth labels~\cite{hendrycks2019using}.
	In this paper, we extend our previous work, based on path integral clustering (PIC) \cite{singh2021pic,Singh2020}, with metric learning framework inspired by neural PLDA  \cite{Ramoji2020}. 
	 The previous work explored graph based clustering with self-supervised representation learning. The learned representations were used to derive cosine affinity score based adjacency matrix. This adjacency matrix, containing pair-wise similarity scores, was used in path integral clustering for speaker diarization~\cite{singh2021pic}. In the proposed work, we further explore a learnable metric based on neural PLDA in the self-supervised learning framework. In particular, both the embeddings and the adjacency matrix for graph based clustering are jointly learned. Using this joint learning, we show significant performance improvements over baseline systems and previous models based on self-supervised graph clustering methods~\cite{Singh2020,singh2021pic}.
	 
	\section{Related Work And Contributions}\label{sec:relatedWork}
	The most common clustering approach used in speaker diarization is based on agglomerative hierarchical clustering (AHC)~\cite{day1984efficient}. 
	The affinity measures explored for AHC consists of cosine similarity score~\cite{silovsky2012speaker} or  PLDA \cite{ioffe2006probabilistic, sell2014speaker}. Other methods for clustering include k-means  \cite{shum2011exploiting} and spectral clustering~\cite{wang2018speaker}.  The  long short term memory (LSTM) network used for affinity computation, proposed by Lin et. al.~\cite{Lin2019} and a fully supervised speaker diarization,  using unbounded interleaved recurrent neural networks (RNN), proposed by Zhang et. al.~\cite{zhang2019fully}, have been  investigated recently. 
	

 The approaches based on re-segmentation~\cite{diez2018speaker,singh2019leap} consists of using the clustering based results to initialize a hidden Markov model (HMM) based clustering. Recently, the x-vector embedding based HMM variational Bayes, termed as VBx, has shown promising results~\cite{landini2020bayesian}. 
 

For self-supervised clustering, the loss functions based on k-means  \cite{yang2017towards}, spectral clustering loss~\cite{shaham2018spectralnet} and agglomerative clustering~\cite{yang2016joint} have been investigated for image/text. A recent work based on neural modeling of discriminatively trained PLDA model \cite{Ramoji2020} was proposed for speaker verification. The neural formulation of PLDA allows the learning of the parameters in a Siamese network. 

This paper extends our previous works on self-supervised learning and graph-based clustering \cite{Singh2020,singh2021pic}. The previous works proposed  representation learning and graph based clustering in an iterative  self-supervised learning framework. The triplet training loss based on cosine similarity was used in the representation learning. The final embeddings were used with the agglomerative hierarchical clustering (AHC)~\cite{Singh2020} or with a more robust graph based clustering called path integral clustering (PIC) \cite{singh2021pic}. The affinity measures in both the approaches was based on the cosine similarity score. 

In this work, we include the PLDA parameters as a learnable metric in the self-supervised learning framework  inspired by  Ramoji et. al~\cite{Ramoji2020}. The PLDA parameters are learnable along with the neural network parameters for the embedding extraction. The entire model is trained using binary cross entropy loss (BCE). The advantage of the proposed approach over the previous work is the direct optimization of the adjacency matrix used in the graph based clustering. We call the proposed approach as self-supervised PLDA based metric learning with path integral clustering (SelfSup-PLDA-PIC).

	
	\section{Background}
	This section describes the pre-processing steps, the probabilistic linear discriminant analysis (PLDA) model and the path integral clustering algorithm used in our approach.
	

	\begin{figure*}[t!]
	\centering 
	\includegraphics[width=15cm]{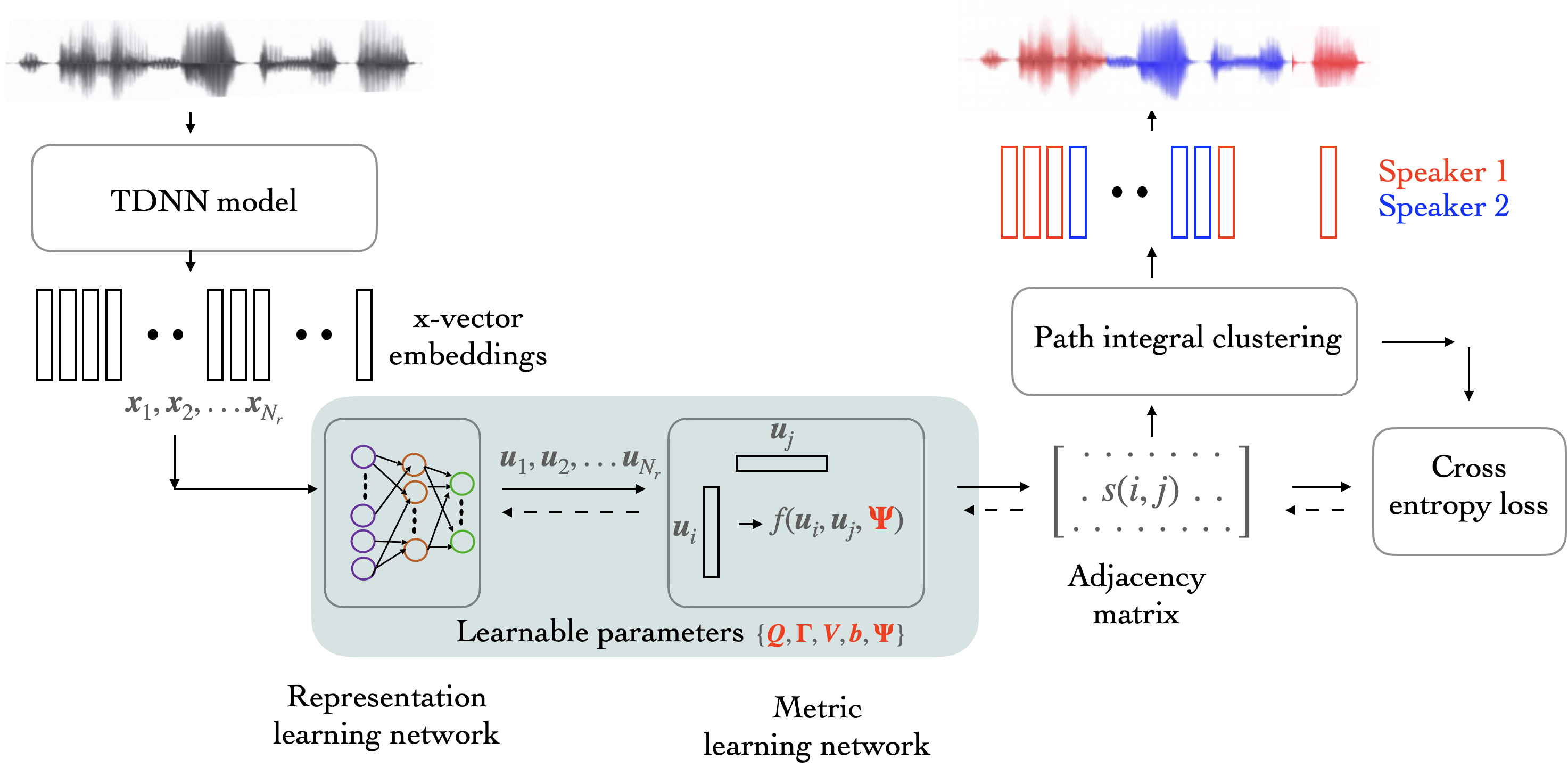}
	\caption{Block schematic of the proposed self-supervised metric learning approach to speaker diarization. }
	\label{fig:blockSchematic}
	\end{figure*}

	\subsection{Pre-processing steps}\label{sec:preprocess}
The x-vectors of each segment in a given recording, extracted using the TDNN network~\cite{snyder2018x}, are mean normalized and whitened. The whitened x-vector features are then processed using length normalization \cite{garcia2011analysis}. Further, a principal component analysis (PCA) based  dimensionality reduction at the recording level is also applied  \cite{zhu2016online}. 
	
\subsection {PLDA model}
	The PLDA is a generative model which factorizes the input into the speaker factor and the channel factor.
	The simplified and widely used model is a generative linear-Gaussian model \cite{ioffe2006probabilistic}, where x-vector $\x\in \mathbb{R}^D$ represents segment  embedding. The vector $\y\in \mathbb{R}^D$ represents the speaker factor. The distribution of $\x$ given the speaker factor $y$ is assumed to be a Gaussian distribution,
	\begin{equation}
		p(\x|\y) = N(\x;\y,\bm{\Phi_w}) 
	\end{equation}
	where $\bm{\Phi_w}\in \mathbb{R}^{DXD}$ is the within class covariance matrix. The latent vector $\y$ is assumed to be distributed according to prior distribution:
	\begin{equation}
		p(\y) = N(\y;\m,\bm{\Phi_b})
	\end{equation}
	where $\m \in \mathbb{R}^D$ and $\bm{\Phi_b}\in \mathbb{R}^{DXD}$ are global mean and between-speaker covariance matrix respectively. Further, $\bm{\Phi_w}$ and $\bm{\Phi_b}$ can be simultaneously diagonalized using diagonalizing transform $\bm{V}\in \mathbb{R}^{DXD}$:
	\begin{eqnarray}\label{eq:diagonalize}
		\V\bm{\Phi_w}\V^T=\bm{I},~~
		\V\bm{\Phi_b}\V^T=\bm{\Psi} 
	\end{eqnarray}
	where $\bm{\Psi}$ is a diagonal covariance matrix. Given this model, the generative graphical model can be expressed in terms of $\bm{u},\bm{v}$, {\color{black}where $\bm{u},\bm{v}$  denote the latent vectors} representing speaker variable and the speaker embedding in the projected space respectively. If $\bm{A}=\V^{-1}$, the generative  model is expressed as:
	\begin{eqnarray}\label{eq:genModel}
	p(\boldsymbol{v}) &=& N(.|\bm{0},\bm{\Psi}) \\  \nonumber
	p(\boldsymbol{u}|\boldsymbol{v}) & =& 
	N(.|\boldsymbol{v},\bm{I}), \\ 
	\x&=&\m+\bm{A}\bm{u}  \nonumber
	\end{eqnarray}
\subsubsection{PLDA Score computation}
\label{sec:pldascoring}
	The embeddings from the PLDA model are used to obtain the pair-wise similarity score matrix $\boldsymbol{S}$ which captures the similarity between embeddings in the speaker space. The similarity score between a pair of embeddings $\x_i$ and $\x_j$, denoted as $s(i,j)$, is based on the log-likelihood ratio score between  same-speaker hypothesis $\mathcal{H}_s$ and different-speaker hypothesis $\mathcal{H}_d$. This can be computed using the PLDA model.
	We project the embeddings $\x$ into latent space using Equation (\ref{eq:genModel}), 	\begin{equation}\label{eq:projected}
		\bm{u}=\bm{A}^{-1}(\x-\m)=\V(\x-\m)=\V\x-\bm{b}
	\end{equation}
	The similarity score can be computed as \cite{ioffe2006probabilistic}:  
	\begin{eqnarray}\label{eq:plda_score}
		s(i,j)&=-\frac{1}{2}\sum_{k=1}^{d}log(\bm{\Psi}[k]+\frac{1}{2})+2log(\bm{\Psi}[k]+1)\\\nonumber
		&+\frac{(\Bar {\bm{u}}[k])^2}{\bm{\Psi}[k]+\frac{1}{2}}+\sum_{l\in\{i,j\}}\left((\bm{u}_l[k]-\Bar {\bm{u}}[k])^2+\frac{(\bm{u}_l[k])^2}{\bm{\Psi}[k]+1}\right)
	\end{eqnarray}
	where $\bm{\Psi}[k]$ is the diagonal element of the k-th row of $\Psi$, $\bm{u}_i[k]$ is the k-th dimension of $\bm{u}_i$ and  $\bar{\bm{u}}[k]=\frac{\bm{u}_i[k]+\bm{u}_j[k]}{2}$ 
	
	\begin{algorithm}[t!]
    \SetAlgoLined
    \textbf{Initialize:} Construct a  graph $G=(V,E)$ where,
     vertices $V$ are the input data $\bm{X}=\{x_1,x_2,...,x_{N_r}\}$. The weighted adjacency matrix $\bm{W}$ is computed and the transition probability matrix $\bm{P}$ is obtained by normalizing $\bm{W}$;\\
     Form $n_c$ initial clusters $\mc=\{\mc_1,...,\mc_{n_c}\}$ by assigning each sample $x_i$ to a cluster, using nearest neighbor merging; $N^*=~$required number of speakers \\
     \While{$n_c>N^*$}{
      \begin{enumerate}
          \item Merge $\mathcal{C}_a$ and $\mathcal{C}_b$, if
          $\{\mathcal{C}_a,\mathcal{C}_b\}=\argmax\limits_{\mc_a,\mc_b\in \mc} \mathcal{A}(\mc_a,\mc_b)$\\ where,
          $ \mathcal{A}(\mc_a,\mc_b)$ is given in Equation (\ref{eq:affnty_pic})
          \item $\mc_c\leftarrow\{\mc_c\backslash\{\mc_a,\mc_b\}\cup\{\mc_a\cup\mc_b\}\}$ and $n_c=n_c-1$
          \item Recompute $\mathcal{A}$
      \end{enumerate}
     }
     \textbf{Termination:}\vspace{0.5em} $\mc_c$
     \caption{Path Integral Clustering (Sec. \ref{sec:pic})}
    \label{algo:pic}
    \end{algorithm}
	\subsection{The path integral clustering}\label{sec:pic}
	The path integral clustering (PIC)~\cite{zhang2013agglomerative} is a graph-based agglomerative clustering algorithm, introduced in \cite{singh2021pic} for speaker diarization.
	
	In PIC, a directed graph is created such that vertices represent the input features connected by the set of edges. 
	 
 	Let the x-vector embeddings from a recording $r$ be denoted as $\bm{X}=\{\x_1,\x_2,...,\x_{N_r}\}\in \mathbb{R}^D$, where $N_r$ is the total number of embeddings present. 
 	We compute an adjacency matrix $\bm{W}$ using the similarity scores (cosine or PLDA scores). From the adjacency matrix, we select only the $K$-nearest neighbors of $\bm{x}_i$. The matrix $\bm{W}$ is converted to transition probability matrix $\bm{P}$ by dividing each row with its row sum.

	The PIC involves computation of path integral $\mathcal{S}_{\mathcal{C}_a}$ and conditional path integral $\mathcal{S}_{\mathcal{C}_{a} \mid \mathcal{C}_{a} \cup \mathcal{C}_{b}}$ for every cluster pair $\mathcal{C}_a$ and $\mathcal{C}_b$ at each step of merging as follows:
	\begin{eqnarray}
		\mathcal{S}_{\mathcal{C}_a} &=& \frac{1}{|\mathcal{C}_a|^2}\bm{1}^T\left(\bm{I}-\sigma\bm{P}_{\mathcal{C}_a}\right)^{-1}\bm{1} \label{eq:pic_ca}\\
		\mathcal{S}_{\mathcal{C}_{a} \mid \mathcal{C}_{a} \cup \mathcal{C}_{b}} &=& \frac{1}{|\mathcal{C}_a|^2}\bm{1}_{\mathcal{C}_a}^T\left(\boldsymbol{I}-\sigma\bm{P}_{\mathcal{C}_{a} \cup \mathcal{C}_{b}}\right)^{-1}\bm{1}_{\mathcal{C}_{a}} \label{eq:pic_ca_cb}
	\end{eqnarray}
	where, $|\mathcal{C}_a|$, cardinality of $\mathcal{C}_a$, is used for normalization of the path integrals. $\bm{P}_{\mathcal{C}_a}$ and $\bm{P}_{\mathcal{C}_{a} \cup \mathcal{C}_{b}}$ are the sub-matrices of the transition probability matrix $\bm{P}$,  column vector $\bm{1}$ is a vector of all ones and of size $|\mathcal{C}_a|$ and $\bm{1}_{\mathcal{C}_a}\in \mathbb{R}^{|\mathcal{C}_a \cup \mathcal{C}_b|}$ is a binary column vector containing ones and zeros corresponding to the nodes of  $\mathcal{C}_a$  and $\mathcal{C}_b$ respectively. The scalar $0<\sigma<1$ introduces discounting of longer paths. 
	The cluster affinity measure for the PIC algorithm is computed as,
	\begin{equation}\label{eq:affnty_pic}
		\mathcal{A}\left(\mathcal{C}_a,\mathcal{C}_b\right)= \mathcal{S}_{\mathcal{C}_{a} \mid \mathcal{C}_{a} \cup \mathcal{C}_{b}}-\mathcal{S}_{\mathcal{C}_{a}}  + \mathcal{S}_{\mathcal{C}_{b} \mid \mathcal{C}_{a} \cup \mathcal{C}_{b}}-\mathcal{S}_{\mathcal{C}_{b}} 
	\end{equation}
	where, $\mathcal{S}_{\mathcal{C}_{a} \mid \mathcal{C}_{a} \cup \mathcal{C}_{b}}-\mathcal{S}_{\mathcal{C}_{a}}$ is the incremental path integral of $\mathcal{C}_{a}$. It represents the sum of weighted paths between $\mathcal{C}_{a}$ and  $\mathcal{C}_{b}$ such that the starting and ending vertices are in $\mathcal{C}_{a}$. Thus, higher affinity shows denser connections between clusters. The clusters with maximum affinity are merged at each time step. The pseudocode is given in Algorithm \ref{algo:pic}.

	\section{Proposed Approach}\label{sec:proposedWork}
	The block schematic of the proposed self-supervised metric learning with graph based clustering algorithm (SelfSup-PLDA-PIC) is given in Figure \ref{fig:blockSchematic}. 
 The model consists of a representation learning network and a metric learning network. The x-vectors extracted from short overlapping audio segments are used as inputs to the model.
	The model generates adjacency matrix which is used in PIC.
	
	The SelfSup-PLDA-PIC\footnote[1]{github code link: \url{https://github.com/iiscleap/SelfSup_PLDA.git}} jointly performs representation learning and metric learning using the initial clustering results. The output of clustering generates speaker labels. These labels are used to form same speaker and different speaker score level targets  for the adjacency matrix. The model training is performed using the binary cross entropy (BCE) loss using the target adjacency matrix. {\color{black}Since the model  updates the parameters based on the unsupervised clustering labels, it is known as self-supervised training. Our previous work involved self-supervised representation learning using fixed cosine similarity scoring~\cite{Singh2020} whereas, here we have introduced metric learning block inspired by PLDA scoring (Sec. \ref{sec:pldascoring}). The similarity score is computed using Equation (\ref{eq:plda_score}),  where, along with embeddings $\bm{u}$, the PLDA parameters  $\bm{\Psi}$ is also learned. Therefore, it is also referred as neural PLDA~\cite{Ramoji2020}.}
In the following sub-section, we discuss model architecture and the joint representation learning and metric learning approach.
	\subsection{Model architecture and training}
	The representation learning network is a three layer DNN with $\{D,d,d\}$ units where $D$ is the x-vector dimension.
	 It takes x-vectors of a recording $r$ as input and generates d-dimensional embeddings $\bm{u}=\{\bm{u}_1,...,\bm{u}_{N_r}\}$. Let,  $\{\bm{Q},\bm{\Gamma},\bm{V}\}$ denote the learnable weights of each layer respectively and let $\bm{b}$ denote the bias of last layer. The embeddings $\bm{u}$ are passed to the metric learning network, which performs pairwise neural PLDA scoring (discussed in Sec. \ref{sec:pldascoring}) using learnable parameter $\bm{\Psi}\in \mathbb{R}^{d \times d}$. In the forward pass, we generate adjacency matrix $\bm{W}$ using pairwise PLDA scores and perform graph based path integral clustering (discussed in Sec.~\ref{sec:pic}). 

	In the backward pass, we compute the ideal adjacency matrix using the clustering solution from the PIC step. The ideal adjacency matrix is a binary matrix consisting of target (label of $1$) and non-target speaker similarity (label of $0$) scores. Using the target and model output, a binary cross entropy (BCE) based loss function is used to update the learnable parameters $\{\bm{Q},\bm{\Gamma},\bm{V},\bm{b},\bm{\Psi}\}$.   A sigmoid non-linearity is also applied on the neural PLDA scores before the BCE loss computation. 

	\section{Experiments}\label{sec:experiment}
	\subsection{Evaluation data}
	\begin{itemize}[leftmargin=*]
		\item \textbf{AMI}: The AMI dataset~\cite{mccowan2005ami} contains meeting recordings from four different sites (Edinburgh, Idiap, TNO, Brno). The ofﬁcial speech recognition partition of AMI dataset comprises of development and evaluation sets consisting of $18$ and $16$ recordings sampled at  $16$kHz respectively.
	We use  the single distant microphone (SDM) condition  of AMI dataset for experiments. We also compare results on beamformed multi-distant microphone (MDM) recordings with other published results.
	The number of speakers and duration ranges from 3-5 and $20$-$60$ mins respectively. For the AMI dataset experiments, we use the diarization error rate (DER) metric with a $250$ms collar and by ignoring the overlap regions (as is the common practice on the AMI dataset). 
		
		\item \textbf{DIHARD III}: The third DIHARD challenge dataset \cite{ryant2020dihard} was released as the third in series of DIHARD speech diarization challenges. It consists of development and evaluation sets of recordings with duration of $0.5$-$10$ mins. These recordings are drawn from $11$ domains including audio-books, telephone recordings, clinical interviews, restaurant conversations, web videos etc.  The number of speakers varies from $1$-$10$ with diverse regions of overlapping speech and speaker turn behavior. 
		There are $254$ and $259$ recordings in the development and evaluation sets respectively. For the DIHARD dataset experiments, we use the DER metric with the overlaps and without providing a collar region. 
	\end{itemize}
	\subsection{Baseline Model}
	Our baseline model is based on DIHARD III baseline recipe \cite{ryant2020dihard}. It involves feature extraction followed by x-vector embedding extraction. The x-vectors are extracted using the extended-TDNN (ETDNN) \cite{snyder2019speaker} network. For training the ETDNN model, we use $40$D mel-filterbank features using a $25$ms window with $10$ms shift. 
	
	The $13$-layer ETDNN model follows the architecture described in \cite{snyder2019speaker,zeinali2019but}. The ETDNN model is trained on the VoxCeleb1 \cite{nagrani2017voxceleb} and VoxCeleb2 \cite{Chung2018} datasets, for speaker identification task, to discriminate among the $7,146$ speakers. It has $4$ TDNN layers which alternates with $4$ fully connected layers of size $1024$D. This is  followed by $2$ feed forward layers containing $\{1024,2000\}$ units. The segment pooling layer is of  $4000$D, containing mean and standard deviation computed at the segment level for the $1500$D layer. From the segment level features, the $512$ dimensional output of the affine component from the $11^{th}$ layer is taken as the x-vector embedding. The pre-processing steps mentioned in section \ref{sec:preprocess} are applied to x-vectors which include whitening transform obtained from the DIHARD development set, length normalization and recording level PCA.
	We preserve $30$ dimensions for the AMI dataset. For the DIHARD dataset, we choose the number of dimensions which preserves $30$\% of total variance. 
	
	The PLDA model is trained using $3$sec x-vectors extracted from the subset of Voxceleb-1  and 2. These x-vectors are whitened using a PCA transform learned from DIHARD development set. We use the same PLDA model for the DIHARD and AMI datasets. 	However, we extract the embeddings using a segment size of $1.5$~s and a temporal shift of $0.75$~s for the AMI dataset due to longer duration recordings while the segments of size $1.5$~s are extracted with a shift of $0.25$~s for the DIHARD dataset.
		
	\begin{figure*}[t!]
        \centering
        \includegraphics[trim={2.8cm 2.1cm 0.5cm 2.4cm},clip,width=\linewidth]{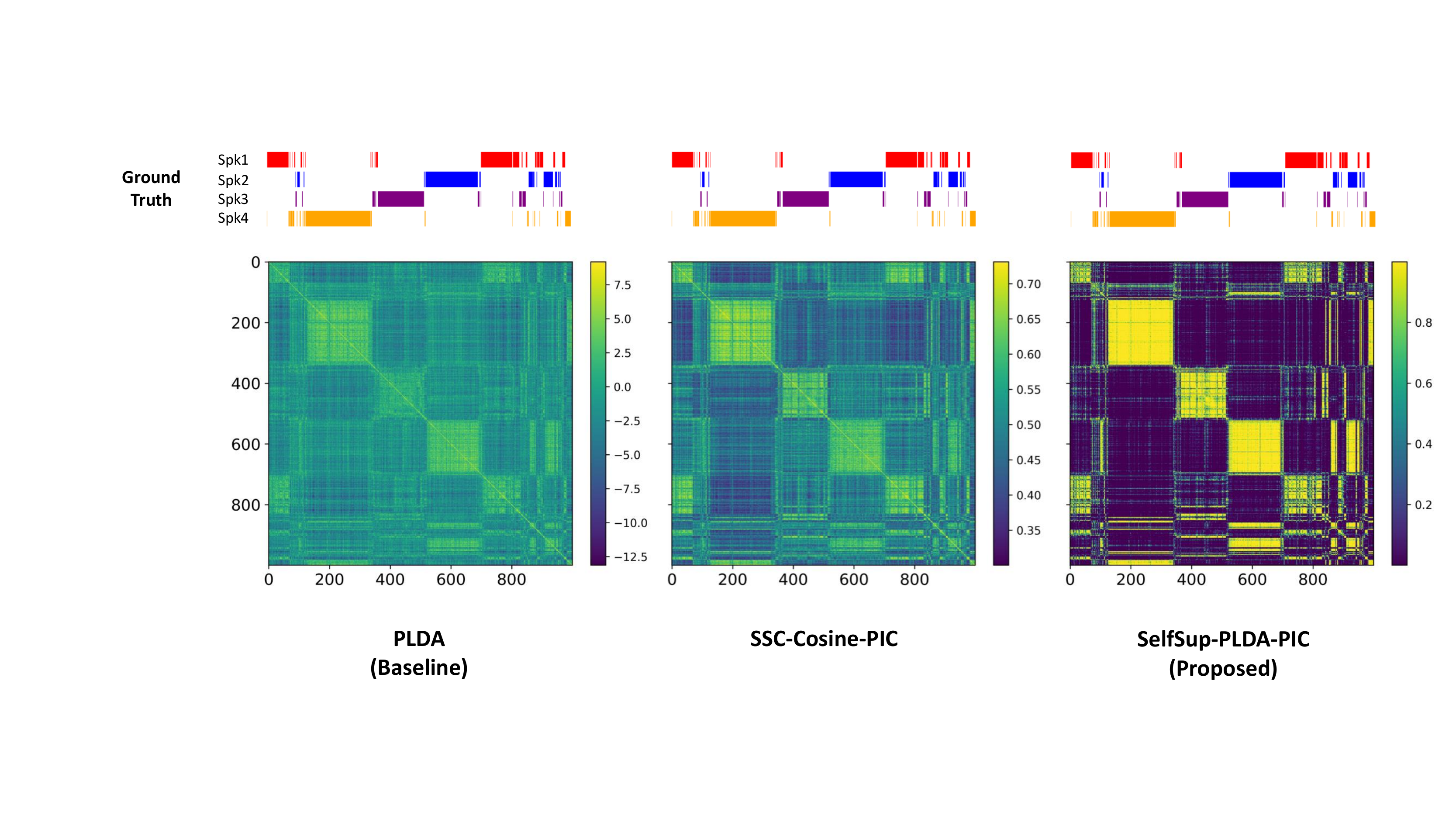}
        \vspace{-1.2cm}
       \caption{Similarity score matrices using PLDA, SSC-Cosine-PIC \cite{singh2021pic} and
       SelfSup-PLDA-PIC (proposed) for a 4-speaker recording from AMI development set. The ground truth labels are plotted across time on top of affinity matrices for comparison.}
        \label{fig:score_comparision}
    \end{figure*}

\subsection{Model Initialization}
The initialization is a critical step for self-supervised training to generate reliable labels. We initialize our model parameters $\{\bm{Q},\bm{\Gamma}\}$ using the whitening transform and the recording level PCA from the baseline system respectively. The third layer's weight and bias $\{\bm{V},\bm{b}\}$ of representation learning network are initialized with diagonalizing tranform $\V$ and bias $\bm{b}$ from PLDA Equation (\ref{eq:projected}),  obtained after applying recording level PCA transform. Similarly, we initialize metric learning parameter $\bm{\Psi}$ using diagonal between class covariance matrix defined in Equation (\ref{eq:diagonalize}).

We perform initial clustering (AHC/PIC) till initial $N^0$ number of clusters.   The value of $N^0$ is based on a threshold applied to the similarity scores for the AHC system. With the PIC, we use the stopping threshold applied on eigen-values of PIC affinity matrix (Equation (\ref{eq:affnty_pic})) \cite{singh2021pic} to estimate $N^0$. We select a threshold higher than the optimal threshold to avoid over-clustering. 
	
For the AMI dataset, the AHC threshold is set as $th=0.0$ to obtain $N^0$ for SelfSup-PLDA-AHC training. For the SelfSup-PLDA-PIC system,  the eigen-value based threshold is set at $th=0.7$ for initial clustering in the self-supervised training. For the  DIHARD dataset, the threshold for the SelfSup-PLDA-AHC system is set at $th=-0.7$. Since the DIHARD dataset has huge variation in number of speakers ($1$-$10$), the number of speakers estimated by AHC is used as $N^0$ in SelfSup-PLDA-PIC experiments.  
	  
\subsection {Choice of hyper parameters}
The hyper-parameters involved in our approach are selected based on the performance on the development set of both datasets. The nearest neighbour $K$ and scaling factor $\sigma$ are the hyper-parameters for PIC. 
The value of $K=30$ and $\sigma=0.1$ are the best values obtained for the AMI dataset. Similarly for DIHARD dataset, we found the best values of $K=40$ and $\sigma=0.5$. After model training,  a temporal continuity of similarity scores can be incorporated \cite{singh2021pic}. This is done by multiplying the similarity score  $s({i,j})$ with an exponential decay given by,
 	\begin{equation} 
 		s'(i,j) = s(i,j)\beta^{min(n_b,|i-j|)}
 		\label{eq:temp_continuity}
 	\end{equation}
where, $\beta$ is a  positive decay factor $<1$,   $|i-j|$ is the absolute segment index difference value of embeddings from the  $i$th and $j$th segment, and $n_b$ is the maximum value of the decay constant $\beta$.


\section{Results}

	\subsection{AMI dataset}
The results for various system configurations with the AMI development and evaluation datasets are reported in Table~\ref{tab:ami_known}. These experiments use the ETDNN based x-vectors. We consider two cases in evaluation, with the known number of speakers $N^*$ and with the unknown number of speakers. The baseline system is the x-vector with PLDA scoring and AHC. The use of graph based clustering with PIC improves the baseline system significantly. 

The self-supervised clustering (SSC) with cosine based affinity matrix proposed previously \cite{singh2021pic} further improves over the PIC based system.  The joint metric learning with the representation learning, proposed in this paper, denoted as SelfSup-PLDA, is shown to provide significant improvements over the previously proposed SSC-Cosine model.  
The relative DER improvement over the baseline system for the SelfSup-PLDA with temporal continuity is $66$\% and $60$\% for the AMI development and evaluation datasets for the condition with unknown number of speakers.  Further, the application of VBx re-segmentation~\cite{landini2020but} on the outputs from the SelfSup-PLDA-PIC system provides final DER values of $2.9$\% and $4.2$\% on the development and evaluation datasets. Our VBx setup is based on baseline ETDNN x-vectors with PLDA adapted from DIHARD dev set. To the best of our knowledge, these results on the SDM data of the AMI corpus constitute the lowest DER reported in literature.

The results using residual network (ResNet101) based x-vector embeddings \cite{landini2020bayesian} are shown in Table~\ref{tab:ami_unknown}.  For these experiments, we use the ResNet 101 architecture. The first layer is a 2D convolutional layer. This is followed by four residual blocks. Each residual block consists of $3$ residual convolutions. The output of the residual blocks is fed to the statistics pooling layer (from each of the $8$ heads). The dense network layer following the pooling layer is used to extract the ResNet x-vectors. The training data and the cost function used to train the ResNet model is similar to the ETDNN framework.   

Comparing the system with PLDA scoring and PIC for the ETDNN x-vectors (Table~\ref{tab:ami_known}) and the ResNet x-vectors (Table~\ref{tab:ami_unknown}), we find that the ResNet x-vectors improve significantly over the ETDNN based x-vectors although both models contain similar no. of parameters ($\sim 10^6$). The performance on the evaluation data for this system is $6.2$\% DER. Further, even with this improved baseline model, the proposed approach of SelfSup-PLDA with graph based clustering and the incorporation of the temporal continuity yields significant improvements. For the self supervised metric learning, it is seen that the final results from either of the x-vector models achieve similar DER results on the AMI evaluation dataset.

\subsection{Adjacency matrix analysis} 

The similarity score matrix (adjacency matrix used in graph clustering) for the baseline x-vector PLDA system (left), self-supervised representation learning with cosine scoring~\cite{singh2021pic} (middle) and the proposed self-supervised metric learning (right) are shown in Figure~\ref{fig:score_comparision}. The adjacency score matrix used in the proposed approach is processed with sigmoid non-linearity for training with BCE loss. The ground truth speaker activity for the four speakers in this recording is also shown in this figure. The same speaker regions of the similarity matrix with the baseline x-vector PLDA system are not well pronounced. The self-supervised embedding learning with cosine scoring improves the contrast between the scores from same speaker and across speaker segments. The proposed approach of metric learning with self-supervised principles is seen to provide the best contrast between the scores from same speaker and across speaker regions of the given audio recording. This increase in contrast partly explains the significantly improved DER results observed in Table~\ref{tab:ami_known} for the proposed SelfSup-PLDA+PIC model. 

\begin{table}[t!]
		\caption{\color{black}{DER (\%) using ETDNN x-vectors on the AMI dataset.}}
		\vspace{0.2cm} 
		\label{tab:ami_known}
		\centering
		\resizebox{\columnwidth}{!}{\begin{tabular}{|l|c|c|c|c|} 
				\hline
				
				\multirow{2}{*}{\textbf{System}} & \multicolumn{2}{c|}{\textbf{Known $N^*$}}& \multicolumn{2}{c|}{\textbf{Unknown $N^*$}} \\ 
				\cline{2-5}
				& Dev.   & Eval. & Dev. & Eval.  \\ 
				\hline\hline
				
				x-vec + PLDA + AHC   &15.9  & 12.2 &13.1  &12.3   \\ 
				x-vec + PLDA + PIC   & \color{black} 5.1   & \color{black}10.2 & \color{black}5.8 & \color{black}11.4                    \\ 
				SSC-Cosine-PIC \cite{singh2021pic}  &5.3 &6.2 &6.5 &8.4 \\ \midrule 
				SelfSup-PLDA-AHC   & 7.9 & 7.3     &  7.7    & 9.4              \\
				 
				SelfSup-PLDA-PIC  & \color{black} 4.2 & \color{black} 6.2  & \color{black} \textbf{4.4} & \color{black} 6.9                     \\ 
				+ Temporal continuity & {\color{black}\textbf{4.2}}  & {\color{black}\textbf{4.2}}    & {\color{black}\textbf{4.4}} & {\color{black}\textbf{4.9}}            \\
					SelfSup-PLDA-PIC + VBx~\cite{landini2020but}  & -  &-   & \textbf{\color{black}2.9} & \textbf{\color{black}4.2}            \\
				\hline
		\end{tabular}}
		\vspace{-0.4cm}
	\end{table}

\begin{table}[t!]
		\caption{\color{black}{DER (\%) using the Resnet x-vectors on the AMI dataset.}}
		\vspace{0.2cm}
		\label{tab:ami_unknown}
		\centering
		\resizebox{0.75\columnwidth}{!}{\begin{tabular}{|l|c|c|} 
				\hline
				
				\multirow{2}{*}{\textbf{System}} &  \multicolumn{2}{c|}{\textbf{Unknown $N^*$}} \\ 
				\cline{2-3}
				& Dev. & Eval.  \\ 
				\hline\hline

				x-vec + PLDA + PIC   & 6.0 & 6.2             \\
			
				
				SelfSup-PLDA-PIC   &4.6  &  6.0                   \\ 
				+ Temporal continuity  &4.4  & \textbf{4.3}             \\
				SelfSup-PLDA-PIC + VBx~\cite{landini2020but}   & \textbf{3.4}  & 4.5            \\
				\hline
		\end{tabular}}
		\vspace{-0.5cm}
	\end{table}	
	
\begin{table}[t!]
		\caption{\color{black}{DER (\%) on the MDM recordings of AMI dataset (without TNO recordings).}}
		\vspace{0.2cm}
		\label{tab:ami_comparison}
		\centering
		\resizebox{0.85\columnwidth}{!}{\begin{tabular}{|l|c|c|} 
				\hline
				
				\multirow{2}{*}{\textbf{System}} &  \multicolumn{2}{c|}{\textbf{Unknown $N^*$}} \\ 
				\cline{2-3}
				& Dev. & Eval.  \\ 
				\hline\hline
				x-vec(ResNet101)+AHC+VBx \cite{landini2020bayesian}   & 2.78  & 3.09             \\
				ECAPA-TDNN \cite{dawalatabad2021ecapa}	& 3.66 & \textbf{3.01} \\
				
				x-vec(ETDNN)+ SelfSup-PLDA-PIC   &5.38  &4.63 \\ 
				-- + VBx~\cite{landini2020but}    &\textbf{2.18}  &3.27 \\
				\hline
		\end{tabular}}
	\end{table}	
\subsection{Comparison with prior literature}
We attempt to compare the recent works proposed in Landini et. al. \cite{landini2020bayesian} and Dawalatabad et. al.~\cite{dawalatabad2021ecapa} with the work proposed in this paper. These previous works report using the beamformed audio from the AMI corpus (multi-distant microphone or MDM), while all the previous results reported in this work used the more challenging single distant microphone (SDM) condition. In order to make the direct  comparison, we did not perform any adaptation or fine-tuning on the MDM data. Rather, the same models used for the SDM evaluations are employed to perform diarization on the MDM recordings. Further, keeping in line with the prior works, we have also omitted the TNO recordings in the development and evaluation set in these results. The comparative analysis is shown in Table~\ref{tab:ami_comparison}. The addition of VBx based re-segmentation to the proposed approach provides the best performance on the AMI development set compared to the prior works, with a final DER of $2.18$\%. Further, the result on the evaluation set (DER of $3.27$\%) is slightly inferior to the best reported result of $3.01$\%. This analysis highlights that, even without fine-tuning or adapting the self-supervised model parameters to the MDM condition, the techniques reported in this paper can match the best state-of-art results for the beamformed audio recordings.    

\subsection{DIHARD dataset}
The results on the DIHARD dataset are reported in Table~\ref{tab:dihard}. The PIC approach improves slightly over the AHC approach used in the baseline system~\cite{ryant2020third}. The self-supervised learning approach with cosine similarity \cite{singh2021pic} degraded the performance over the baseline system. This was analyzed to be partly due to the reduced duration of the recordings, large number of speakers within the given recording and the lack of robustness in the simple cosine similarity scoring. 

The proposed approach of SelfSup-PLDA improves over both the AHC and PIC systems respectively. Without the VBx re-segmentation, the best results are achieved for the SelfSup-PLDA-AHC model. However, the VBx re-segmentation did not improve over the re-segmentation applied on the baseline model. 

As seen here, the improvements in the DIHARD dataset are less significant compared to those observed in AMI dataset. The primary reason for this reduction in improvement is the shorter duration files ($0.5$-$10$min duration) in DIHARD dataset compared to the $20$-$60$min duration of the AMI recordings. The self-supervised metric learning approaches proposed in this work rely on recording level labels to improve the adjacency matrix used in the graph based clustering. With a reduced number of embeddings, the training of the SelfSup-PLDA model is compromised. Secondly, the DIHARD datasets have diverse domains with some domains having a large number of speakers (more than $7$ speakers) in the given recording.

	\begin{table}[t!]
		\caption{\color{black}{DER (\%) when number of speakers ($N^*$) are unknown for the DIHARD dataset.}}
		\vspace{0.25cm}
		\label{tab:dihard}
		\centering
		\resizebox{0.76\columnwidth}{!}{\begin{tabular}{|l|c|c|} 
				\hline
				
				\multirow{2}{*}{\textbf{System}} & \multicolumn{2}{c|}{\textbf{Unknown $N^*$}} \\ 
				\cline{2-3}
			& Dev. & Eval.  \\ 
				\hline\hline
				
				x-vec + PLDA + AHC \cite{ryant2020third}    &19.7  & 19.5            \\ 
				-- + VBx \cite{landini2020but}  & 17.0 &  16.6          \\
				x-vec + PLDA + PIC    & \color{black} 19.7  & \color{black} 18.9                   \\
				-- + VBx \cite{landini2020but} & \color{black} \textbf{16.8}  & \textbf{16.3}          \\
				SSC-Cosine-PIC \cite{singh2021pic}  &23.9 &21.1 \\ \midrule 
				
				SelfSup-PLDA-AHC    & \textbf{18.9 }     & \textbf{18.2 }            \\
				
				SelfSup-PLDA-PIC & \color{black} 19.2  & \color{black}  \textbf{18.2}                  \\ 
				-- + VBx \cite{landini2020but}  & 17.5  &  17.2         \\
				\hline
		\end{tabular}}
	\end{table}
The large number of speakers also decreases the quality of the pseudo cluster labels used in the self-supervised learning. In order to analyze the impact of the increased number of speakers, we split the results reported in Table~\ref{tab:dihard} into two conditions - recordings having less than or equal to $7$ speakers and those with more than $7$ speakers. This analysis is reported in Table~\ref{tab:dihard-7spk}.  As seen here, the model of self-supervised metric learning provides consistent performance improvements for the recordings having less than $8$ speakers. On the other hand, for recordings with more than $7$ speakers, the self-supervised metric learning results in a performance degradation. As hypothesized earlier, the degradation is attributed to the errors in the psuedo-labels used in the self-supervised learning. In future, a  confidence measure will also be explored in the self-supervised learning framework to avoid over-fitting to the noisy cluster labels. 

	\begin{table}[t!]
		\caption{\color{black}{Average DER (\%) on the DIHARD dataset for recordings with $\leq 7$ speakers and $>7$ speakers }}
		\vspace{0.1cm}
		\label{tab:dihardspks}
		\centering
		\resizebox{0.85\columnwidth}{!}{\begin{tabular}{|l|c|c|c|c|} 
				\hline
				
				\multirow{2}{*}{\textbf{System}} & \multicolumn{2}{c|}{\textbf{$\leq 7$ speakers}}& \multicolumn{2}{c|}{\textbf{$>7$ speakers}} \\ 
				\cline{2-5}
				& Dev.   & Eval. & Dev. & Eval.  \\ 
				\hline\hline
				
				x-vec + PLDA + AHC    &18.0  & 19.3  &36.6  &  27.1          \\ 
				
				x-vec + PLDA + PIC   & \color{black} 17.7  & 17.8  & \textbf{36.5}   & \textbf{24.0}              \\ 
				\midrule 
				
				SelfSup-PLDA-PIC  & \textbf{17.0} & \textbf{17.2} & 39.5 &  28.1                \\ 
			
				\hline
		\end{tabular}}
		\label{tab:dihard-7spk}
	\end{table}
    
\section{Summary}
We have proposed an approach to perform metric learning and clustering jointly for the task of diarization. The metric learning is performed in a self-supervised manner by updating the neural PLDA model using cluster identities provided by graph based path integral clustering. 
Using an iterative procedure of metric learning and clustering, we show that the proposed algorithm provides improved similarity scores and precise speaker clusters. With challenging diarization datasets, we have illustrated the performance improvements obtained using the proposed approach. In particular, the self supervised metric learning algorithm provides the best results reported thus far for the AMI single distant microphone conditions. With the more challenging DIHARD dataset evaluations, the proposed approach did not show improvements when the number of speakers in the given recording was greater than $7$. However, for the recordings with {\color{black}less than $8$ speakers}, the model showed consistent performance improvements over the baseline systems compared. 

\section{Acknowledgements}
The authors would like to thank Abhishek Anand and Michael Free of British Telecom Research for their inputs in the model development. The authors would also like to acknowledge the efforts of Rajat Varma in DIHARD experiments as well as in the manuscript preparation. 

	\bibliographystyle{IEEEbib}
	\bibliography{strings,refs}
	
\end{document}